\documentclass[
reprint,    
nofootinbib, 
aps,
prd,
amsmath,
amssymb,
]{revtex4-1}

\usepackage{graphicx} 
\usepackage{dcolumn}  
\usepackage{bm}       

\usepackage[
colorlinks=true,
citecolor=blue
]{hyperref}
\usepackage{bbm}
\usepackage{lipsum}
\usepackage{amsmath}
\usepackage[detect-all]{siunitx}
\usepackage{appendix}
\usepackage{xcolor}

\definecolor{johnblue}{RGB}{94, 129, 172}

\newcommand{\ket}[1]{| #1 \rangle}
\newcommand{\bra}[1]{\langle #1 |}
\newcommand{\braket}[1]{\langle #1 \rangle}
\newcommand{\ketbra}[2]{\ket{#1}\bra{#2}}

\graphicspath{{img/}} 

\begin{document}

\title{Emergent symmetry in a two-Higgs-doublet model from quantum information and nonstabiliserness}

\date{\today}

\author{Giorgio Busoni}
\email{giorgio.busoni@adelaide.edu.au}
\affiliation{
  University of Adelaide, ARC Centre of Excellence for Dark Matter Particle Physics \& CSSM, Department of Physics, Adelaide SA 5000 Australia
}

\author{John Gargalionis}
\email{john.gargalionis@adelaide.edu.au}
\affiliation{
  University of Adelaide, ARC Centre of Excellence for Dark Matter Particle Physics \& CSSM, Department of Physics, Adelaide SA 5000 Australia
}

\author{Ewan N. V. Wallace}
\email{ewan.n.wallace@adelaide.edu.au}
\affiliation{
  University of Adelaide, ARC Centre of Excellence for Dark Matter Particle Physics \& CSSM, Department of Physics, Adelaide SA 5000 Australia
}

\author{Martin J. White}
\email{martin.white@adelaide.edu.au}
\affiliation{
  University of Adelaide, ARC Centre of Excellence for Dark Matter Particle Physics \& CSSM, Department of Physics, Adelaide SA 5000 Australia
}

\begin{abstract}
Studies of scattering processes in scalar models with two Higgs doublets have recently hinted at a connection between the absence of flavour-space entanglement in \(\Phi^+\Phi^0\) scattering and an emergent \(\mathrm{SO}(8)\) symmetry in the scalar potential. We extend the analysis to all scattering channels with two particles in the external states by treating the process as a four-qubit system in the weak isospin and flavour subspaces of the \(2\)-particle state. We work with a generic quantum information-theoretic principle encoded by the commutativity of the initial state density matrix with the transition matrix (at leading order in perturbation theory). This yields a special case of the entanglement minimisation conditions previously derived in the literature, and we interpret the principle in terms of the conservation of nonstabiliserness (or ``magic''). Working at leading order in the quartic couplings, we find a consistent set of conditions that implies an \(\mathrm{SO}(8)\) symmetry on the quartic part of the potential for scattering an arbitrary initial state, but a smaller \(\mathrm{SU}(2)_R\) symmetry when the initial state is chosen to have definite isospin. This follows by accounting for Bose symmetry in the initial state, which introduces entanglement between the isospin and flavour subspaces.
\end{abstract}

\maketitle

\section{Introduction}

Symmetry plays a crucial role in the formulation of the Standard Model (SM) of particle physics and its proposed extensions. Yet the particular form of the SM gauge group is ultimately determined phenomenologically. The discovery of a firm and consistent principle from which symmetry can arise in quantum field theory would not only be theoretically compelling, but also invaluable in guiding the model builder interested in extending the SM to address its many and well-known problems.

Recent research has suggested that ideas from quantum information theory may be relevant in this context. A general quantum information framework that connects global symmetries to the relativistic $\mathcal{S}$-matrix for \(2 \to 2\) particle scattering was recently presented in Ref.~\cite{McGinnis:2025brt}. Minimisation of entanglement has been shown to lead to emergent symmetries in non-relativistic baryon scattering~\cite{Beane:2018oxh,Low:2021ufv,Liu:2022grf,Liu:2023bnr} and, potentially, to the structure of the Cabibbo-Kobayashi-Maskawa and Pontecorvo-Maki-Nakagawa-Sakata matrices that describe flavour mixing in the quark and lepton sectors~\cite{Thaler:2024anb}. Minimisation of entanglement in the process $\Phi^+_i \Phi^0_j\rightarrow \Phi^+_k \Phi^0_l$ (where $i,j,k,l$ denote different flavours) in a two-Higgs-doublet (2HDM) extension of the Standard Model has been shown to lead to an emergent maximal global symmetry~\cite{Carena:2023vjc}. On the other hand, it has been shown that entanglement minimisation in other allowed scattering processes in the same model does not lead to an emergent symmetry~\cite{Chang:2024wrx,Kowalska:2024kbs}, and indeed imposes conditions inconsistent with those derived from elastic \(\Phi^+ \Phi^0\) scattering. Further, the \textit{maximisation} of entanglement has also been suggested as a symmetry-generating mechanism in the 2HDM~\cite{Carena:2025wyh}, and the same principle has been put forward to explain the structure of the QED vertex~\cite{Cervera-Lierta:2017tdt} and explored in the context of pure \(\mathrm{SU}(N)\) gauge theory~\cite{Nunez:2025dch}. Although superficially these findings complicate the picture, it remains possible that there is indeed some consistent principle that underlies the results; further study is warranted in order to clarify the deeper meaning. 

Our aim here is to elucidate the emergence of symmetry in the 2HDM from scalar \(2 \to 2\) scattering processes, focusing on initial states with two incoming \emph{particles} (i.e., total particle number \(\mathcal{P}=2\)), while accounting for Bose symmetry and \(\mathrm{SU}(2)_L\) invariance. We show that the emergence of the maximal \(\mathrm{SO}(8)\) symmetry in the case of \(\Phi^{+} \Phi^{0}\) scattering can be framed as following from the condition that the initial and final state density matrices are equal up to first order in the quartic couplings of the theory. This is a stronger condition than requiring that the transition matrix \(\mathcal{T}\) have vanishing entanglement power~\cite{Zanardi:2000zz}, and we show that it is equivalent to requiring the conservation of \emph{nonstabiliserness}, colloquially referred to as \emph{magic}~\cite{White:2024nuc}, which characterises those states of a system that confer an advantage in quantum computing. Concretely, our condition is equivalent to the requirement that the initial-state density matrix commutes with \(\mathcal{T}\). 

This constraint also restricts the form of the \(\mathcal{T}\)-matrix to be proportional to the identity gate for an arbitrary initial density matrix. This is similar to the case of minimally entangling \(np\) scattering,\footnote{An important difference is that \(\mathcal{T}\) can also access the \(\mathrm{SWAP}\) gate in \(np\) scattering.} where the condition follows from rotational invariance in \(s\)-wave scattering~\cite{Low:2021ufv,Liu:2022grf}. We note however that Bose symmetry plays an interesting role in restricting the form of \(\rho\) and therefore allowing more flexibility in the structure of the \(\mathcal{T}\)-matrix. We will discuss below that the maximal \(\mathrm{SO}(8)\) symmetry in the 2HDM potential follows from the fact that \(\ket{\Phi^+ \Phi^0}\) is a superposition of the triplet and singlet states of \(\mathrm{SU}(2)_L\) with vanishing projection of the third component of weak isospin.  Our main aim is to extend the study of the emergence of symmetry to all \(2 \to 2\) scattering processes with non-vanishing particle number within the same explicitly isospin-invariant formalism. We find that the scattering of an arbitrary initial state results in an \(\mathrm{SO}(8)\) symmetry acting on the quartic part of the potential. Interestingly, when Bose and \(\mathrm{SU}(2)_L\) symmetry are accounted for in the initial state, consistent conditions imprinting an \(\mathrm{SU}(2)_R\) symmetry onto the scalar potential can be derived after imposing \([\mathcal{T},\rho]=0\) when scattering states of definite weak isospin. As we will see, this analysis complements the existing entanglement results regarding the emergence of the alignment limit (which ensures a SM-like Higgs boson). Recent related work can be found in the study of magic in the neutrino sector in Ref.~\cite{Chernyshev:2024pqy}, in the context of QED in Ref.~\cite{Liu:2025qfl}, and investigations of the magic generated in top quark pair production at the Large Hadron Collider in Refs.~\cite{White:2024nuc,CMS-PAS-TOP-25-001,Aoude:2025jzc}.

This paper is structured as follows. In Sec.~\ref{background_formalism}, we provide a convenient introduction to the concept of nonstabiliserness, a brief review of the salient features of the 2HDM and an explanation of how to treat scalar scattering in flavour-space as a two qubit system. In Sec.~\ref{magic_cons_align}, we compute the initial and final state magic for the process \(\Phi^+\Phi^0\to\Phi^+\Phi^0\) at leading order in perturbation theory, and demonstrate the symmetries that emerge from conservation of magic. In Sec~\ref{sec:general}, we extend our formalism to cover all \(2 \to 2\) scalar scattering processes involving states with total particle number \(\mathcal{P} = 2\) and investigate the resulting emergent symmetries. We conclude in Sec.~\ref{conclusion}.

\section{Background Formalism}
\label{background_formalism}

\subsection{Magic in Quantum Information Theory}
\label{magic_review}

Inspired by recent results for entanglement extremisation, our goal is to consider the viability of alternative or generalised information-theoretic principles in a particle physics setting. Accordingly, we first study another quantum information measure: magic. We follow the presentation of Refs.~\cite{White:2024nuc,White:2020zoz} in outlining the relevant formalism here.

We will be concerned with scattering processes between particles whose flavour information is represented as a two-state quantum system, that is, a qubit, with basis states \(\ket{0},\ket{1}\). For a single qubit, one has the usual Pauli matrices \(\{\mathbbm{1},\sigma_1,\sigma_2,\sigma_3\}\). For a system of \(n\) qubits, one may define the so-called \textit{Pauli strings} by
\begin{equation}
    P=\sigma_{i_1}\otimes\cdots\otimes\sigma_{i_n} \ ,
\end{equation}
where \(i_k=0,1,2,3\). These Pauli strings generate the Pauli \textit{group}, essentially consisting of the Pauli strings weighted by phases of \(\pm 1,\pm i\). Then one may define the \(n\)-qubit \textit{Clifford group} \(\mathbb{C}_n\) consisting of all unitary operators fixing the Pauli group; that is,
\begin{equation}
    \mathbb{C}_n\equiv\left\{U\in\text{U}(2^n):UPU^\dagger=e^{i\theta}P'\right\} \ ,
\end{equation}
where \(P,P'\) are Pauli strings and \(\theta\) is a possible phase. Being the normaliser of the Pauli group in \(\text{U}(2^n)\), this group contains the Pauli strings themselves as well as other such well-known objects as the phase, Hadamard and CNOT gates~\cite{Garcia:2014hia}.

By definition, elements of the Clifford group map eigenstates of Pauli strings to eigenstates of Pauli strings. States that may be reached by acting elements of the Clifford group on computational basis states are called \textit{stabiliser states}; this set is a subset of Pauli-string eigenstates, and includes the computational basis states themselves as well as various others (e.g., the maximally entangled Bell states).

The physical significance of these concepts is this: The \textit{Gottesman-Knill theorem}~\cite{Gottesman:1998hu}, by now a standard result in quantum information theory, asserts that quantum circuits comprised of Clifford gates may be efficiently simulated on a classical computer, even though they may produce highly entangled states. In other words, entanglement alone is insufficient to guarantee computational advantage; an additional resource is required. This resource is referred to as \textit{nonstabiliserness} or \textit{magic}. Magic states are precisely those states that are not stabiliser states.

There exist a number of measures for quantifying magic in the literature. The essential feature of any good magic measure is faithfulness: the measure yields zero if and only if the state is a stabiliser state. The so-called \textit{stabiliser R\'enyi entropies} were first proposed in Ref.~\cite{Leone:2021rzd} as a set of easily-calculated analytic measures. In this work, we use a closely related measure: the \textit{linearised stabiliser entropy}, also introduced in Ref.~\cite{Leone:2021rzd}, defined on a pure state \(\ket{\psi}\) in the \(2^n\)-dimensional qubit space by
\begin{equation}
    M_\text{lin}(\ket{\psi})\equiv1-\frac{1}{2^n}\sum_{P}|\braket{\psi|P|\psi}|^4 \ ,
\end{equation}
where the sum runs over all Pauli strings. That this measure is faithful (and also stable under free operations, i.e., \(M_\text{lin}(U\ket{\psi})=M_\text{lin}(\ket{\psi})\) for all \(U\in\mathcal{C}_n\)) follows from the proof of Theorem 1 in Ref.~\cite{Wang:2023uog}. We will rewrite this definition in terms of the density matrix \(\rho\equiv\ket{\psi}\bra{\psi}\):
\begin{equation}
    \label{magic_formula}
    M_\text{lin}(\rho)=1-\frac{1}{2^n}\sum_{P}\text{Tr}^4(\rho P) \ ,
\end{equation}
by hermiticity of the Pauli matrices (hence also the Pauli strings). Note that this formula expressly assumes a pure state. 

\subsection{2HDM Review}
\label{nhdm_review}
We now turn to the specific 2HDM setting for the \(2\to 2\) scattering we will consider. While inspired by a full two-Higgs-doublet extension of the Standard Model, our construction is deliberately simplified: we are interested only in high-energy scalar scattering and will not concern ourselves with couplings to SM fermions or gauge bosons. Accordingly, we study a theory of two complex scalar fields \(\Phi_i\), each transforming as \((\mathbf{1}, \mathbf{2}, \tfrac{1}{2})\) under the SM gauge group, with hypercharge convention \(Q = I_3 + Y\). In standard notation, the 2HDM potential may be written as
\begin{widetext}
    \begin{align}
        \label{2HDM_gen_potential}
        V=&-m_{11}^2\Phi_1^\dagger \Phi_1-m_{22}^2\Phi_2^\dagger\Phi_2-(m_{12}^2\Phi_1^\dagger\Phi_2+\text{h.c.})+\lambda_1(\Phi_1^\dagger\Phi_1)^2+\lambda_2(\Phi_2^\dagger\Phi_2)^2+\lambda_3(\Phi_1^\dagger\Phi_1)(\Phi_2^\dagger\Phi_2) \nonumber \\
        &+\lambda_4(\Phi_1^\dagger\Phi_2)(\Phi_2^\dagger\Phi_1)+\left(\frac{\lambda_5}{2}(\Phi_1^\dagger\Phi_2)^2+\lambda_6(\Phi_1^\dagger\Phi_1)(\Phi_1^\dagger\Phi_2)+\lambda_7(\Phi_2^\dagger\Phi_2)(\Phi_1^\dagger\Phi_2)+\text{h.c.}\right).
    \end{align}
\end{widetext}
The parameters appearing in the above potential depend on the choice of basis for the doublets, but in our simplified scalar-only scenario, all choices lead to the same physics. For simplicity, we assume explicit CP conservation (so the quartic couplings are real) and a \(\text{U}(1)_\text{EM}\)-invariant minimum.

The alignment limit is defined as the limit in which one of the mass eigenstates of the extended scalar sector assumes the likeness of the 125 GeV SM Higgs; for a review, see Refs.~\cite{BhupalDev:2014bir,Darvishi:2021txa}. In a full 2HDM extension of the SM, the SM Higgs may be identified with the combination \(H_\text{SM}=H\cos{(\alpha-\beta)}+h\sin{(\alpha-\beta)}\) of the two CP-even neutral scalars \(H\) and \(h\), where \(\alpha\) and \(\beta\) diagonalise the CP-even and -odd sectors respectively. (Note that we here assume no spontaneous CP violation; we will see later that this is justified for our analysis.) The cubic couplings of \(H\) and \(h\) to the SM gauge bosons are then modified by factors of \(\cos{(\alpha-\beta)}\) and \(\sin{(\alpha-\beta)}\). It follows that alignment is realised for \(\alpha-\beta\to 0,\pi/2\), when one of \(H\) or \(h\) behaves as the SM Higgs boson; we here assume the case \(\alpha\to\beta\). Ref.~\cite{BhupalDev:2014bir} has demonstrated that this is equivalent to the requirement
\begin{equation}
    \label{align_cond_gen}
        \lambda_7 t_\beta^4-\kappa_2 t_\beta^3+3(\lambda_6-\lambda_7)t_\beta^2+\kappa_1 t_\beta-\lambda_6=0 \ ,
\end{equation}
with \(t_\beta\equiv\tan{\beta}=v_2/v_1\) –– where \(v_i\) is the vacuum expectation value (VEV) of \(\Phi_i\) –– and \(\kappa_i\equiv 2\lambda_i-\lambda_3-\lambda_4-\lambda_5\). Barring a delicate fine-tuning of the quartic couplings, the easiest way to realise natural alignment is to impose a symmetry on the scalar sector that precludes the offending terms. The possible symmetries of the 2HDM potential have been classified~\cite{Darvishi:2019dbh}; only a handful lead to natural alignment, including most notably the maximal \(\text{SO}(8)\) custodial symmetry acting on the eight real degrees of freedom contained in the two doublets. This symmetry results in the potential
\begin{equation}
    V=-m^2\left(\Phi_1^\dagger\Phi_1+\Phi_2^\dagger\Phi_2\right)+\lambda\left(\Phi_1^\dagger\Phi_1+\Phi_2^\dagger\Phi_2\right)^2,
\end{equation}
and must be broken softly in order to avoid generating a largely massless spectrum.

\subsection{Scalar Scattering of Qubits}
\label{scattering}
Here we describe how scalar scattering in the 2HDM setting can be described as a system of two qubits in flavour space. Initially, we will consider only the process \(\Phi^+\Phi^0\to\Phi^+\Phi^0\), previously explored in Ref.~\cite{Carena:2023vjc}. We take the doublet number as our internal flavour degree of freedom. Then we identify our Alice and Bob qubits respectively as \(\ket{i}_A\sim\Phi^+_{i+1}\) and \(\ket{i}_B\sim\Phi^0_{i+1}\), \(i=0,1\). To be even more explicit, we write the doublets as \(\Phi_{ia}\) such that \(i\) enumerates over flavour and \(a\) over the isospin components of the doublet; that is,
\begin{equation}
    \Phi_{i1}=\Phi_i^{+},\quad\Phi_{i2}=\Phi_i^{0}.
\end{equation}
For now, we restrict our attention to flavour, but we will return to isospin later. We thus identify the `Alice' qubit as a linear combination of the \(I_3=+1/2\) components of the doublets and the `Bob' qubit as a combination of the \(I_3=-1/2\) components.

Prior to fixing a basis for the doublets, we may redefine our Higgs fields by an \(\text{SU}(2)\) rotation on the vector \((\Phi_1,\Phi_2)\), mixing flavour in the process. We therefore implicitly fix the basis by rotating the doublets such that \(m_{ij}^2\) is diagonal, i.e., there are no mixed bilinears in the quadratic part of the potential; then the components of the doublets –– our flavour basis states –– are mass eigenstates in the high-energy regime. This rotation has no impact in our toy-model setting without SM couplings.

Since we consider \(2\to 2\) scattering at leading order, the dynamics are restricted to the two-particle Hilbert space \(\mathcal{H}_2\) (a subspace of the full Fock space \(\mathcal{F}\)) and multi-particle states are not accessed perturbatively~\cite{Seki:2014cgq}. A more complete development of the following formalism may be found in Ref.~\cite{Kowalska:2024kbs}; see Refs.~\cite{Thaler:2024anb,Chang:2024wrx} for related approaches. We partition the Hilbert space as \(\mathcal{H}_2\simeq\mathcal{H}_\mathbf{p}\otimes\mathcal{H}_\text{flav}\), where \(\mathcal{H}_\mathbf{p}\simeq L^2(\mathbb{R}^3\otimes\mathbb{R}^3)\) corresponds to the two-particle momentum degrees of freedom and \(\mathcal{H}_\text{flav}\simeq\mathbb{C}^2\otimes\mathbb{C}^2\) corresponds to the discrete two-qubit flavour space. States in these spaces are normalised in the typical way: for the momentum states
\begin{align}
    \label{mom_norm}
    \braket{\mathbf{p}_1\mathbf{p}_2|\mathbf{q}_1\mathbf{q}_2} &= (2\pi)^6 4E_{\mathbf{p}_1} E_{\mathbf{p}_2} \nonumber \\
    &\quad \times\delta^{(3)}(\mathbf{p}_1-\mathbf{q}_1)\delta^{(3)}(\mathbf{p}_2-\mathbf{q}_2) \ ,
\end{align}
and for the flavour
\begin{equation}
    \label{flav_norm}
    \braket{ij|kl}=\delta_{ij}\delta_{kl} \ ,\quad \mathbbm{1}_2=\sum_{ij}\ket{ij}\bra{ij} \ .
\end{equation}
Our pre-scattering initial state is
\begin{equation} \label{eq:product-state-just-flavour}
\begin{aligned}
\ket{\psi_0} &=  \sum_{ij} B_{ij} |ij\rangle \\
&\quad \otimes \int \frac{\mathrm{d}^3 \mathbf{p}_1}{(2\pi)^3 2E_1} \frac{\mathrm{d}^3 \mathbf{p}_2}{(2\pi)^3 2E_2}\phi(\mathbf{p}_1, \mathbf{p}_2) | \mathbf{p}_1 \mathbf{p}_2 \rangle \ ,
\end{aligned}
\end{equation}
where the coefficients \(B_{ij}\) and the momentum wave function \(\phi(\mathbf{p}_1,\mathbf{p}_2)\) satisfy the usual normalisation conditions. The post-scattering final state is then \(\ket{\psi_f}=\mathcal{S}\ket{\psi_0}\), where \(\mathcal{S}\equiv\mathbbm{1}_2+i\mathcal{T}\) is the \(\mathcal{S}\)-matrix and the \(\mathcal{T}\)-matrix elements define the scattering amplitudes \(M_{ijkl}\) by 
\begin{align}
&\braket{\mathbf{p}_3,k;\mathbf{p}_4,l|i\mathcal{T}|\mathbf{p}_1,i;\mathbf{p}_2,j}\equiv iM_{klij}(p_1,p_2\to p_3,p_4) \nonumber \\
    &\times(2\pi)^4\delta^{(4)}(p_1+p_2-p_3-p_4) \ .
\end{align}
We here assume that the flavour and momentum space remain separated, i.e., the post-scattering ket is also a product state in momentum and flavour. Ref.~\cite{Kowalska:2024kbs} has demonstrated that entanglement between these spaces is typically generated in \(2\to 2\) scalar scattering, but it can be shown that ours is a good assumption at leading order in perturbation theory. We wish to study the information-theoretic properties of states in the two-qubit flavour space. Here, the density matrices may be formed as usual:
\begin{equation}
    \rho_0=\ket{\psi_0}\bra{\psi_0}_\text{flav}=\sum_{ijkl}B_{ij}B_{kl}^*\ket{ij}\bra{kl} \ ,
\end{equation}
for the initial-state density matrix, and
\begin{align}
    \label{out_dens_matrix}
    \rho_f&=\mathcal{S}\rho_0\mathcal{S}^\dagger=(\mathbbm{1}_2+i\mathcal{T})\rho_0(\mathbbm{1}_2-i\mathcal{T}) \nonumber \\
    &=\rho_0+i[\mathcal{T},\rho_0]+\mathcal{O}(\lambda^2) \ ,
\end{align}
for the final, where we have used the fact that \(\mathcal{T}=\mathcal{T}^\dagger\) at first order in the perturbation(s) \(\lambda\). Representing the discrete basis as vectors in \(\mathbb{R}^4\), one may now straightforwardly calculate the initial- and final-state magic per Eq.~\eqref{magic_formula}. This is permitted because \(\text{Tr}(\rho_0^2)=1\) by construction, so the initial state is pure; and it may further be shown that \(\text{Tr}(\rho_f^2)=1+\mathcal{O}(\lambda^2)\), so the final state is also pure (to first order in perturbation theory).

\section{Magic Conservation and the Alignment Limit}
\label{magic_cons_align}
We now commence our investigation of conservation of magic and emergent symmetry by computing the magic for the initial and final states of the scattering process \(\Phi^+\Phi^0\to\Phi^+\Phi^0\). As emphasised, we work to lowest order in perturbation theory; at this order, the scattering is facilitated by the contact interactions in the quartic part of the Higgs potential. Referring to Eq.~\eqref{out_dens_matrix}, it is clear that the final-state density matrix will generically have the form \(\rho_f=\rho_0+\mathcal{O}(\lambda)\). Since we assume an arbitrary initial state, \(\rho_0\) is unconstrained; it is thus natural to consider magic \textit{conservation}, in contrast to the entanglement \textit{extremisation} conditions previously explored in the literature.

We enforce conservation of magic by expanding the final-state linearised stabiliser entropy to first order in the lowest-order amplitudes \(M_{ijkl}\); we then subtract the initial-state contribution and demand that the remaining polynomial in \(B_{ij},B^*_{ij}\) vanishes identically. This results in the following condition:
\begin{align}
    \label{cons_of_magic}
    M_{ijkl}=M\delta_{ik}\delta_{jl}
\end{align}
with \(i,j,k,l=0,1\) and where \(M\) is fixed and real. That is, the amplitudes \(M_{ijij}\) are equal and real for all \(i,j\), and all other amplitudes vanish. We stress that this condition is valid for any theory admitting a discrete qubit subspace and satisfying the assumptions of Sec.~\ref{scattering}, not just the 2HDM toy model studied in this work. We also note that in some sense, the condition of Eq.~\eqref{cons_of_magic} is trivial, because we then have
\begin{align}
    \braket{ij|[\mathcal{T},\rho_0]|kl}&=\sum_{mn}\left(M_{ijmn}B_{mn}B^*_{kl}-M_{klmn}^*B_{ij}B^*_{mn}\right) \nonumber \\
    &=\sum_{mn}M\left(\delta_{im}\delta_{jn}B_{mn}B^*_{kl}-\delta_{km}\delta_{ln}B_{ij}B^*_{mn}\right) \nonumber \\
    &=M (B_{ij}B^*_{kl}-B_{ij}B^*_{kl})=0
\end{align}
and hence \(\rho_f=\rho_0+\mathcal{O}(\lambda^2)\). In fact, it is straightforward to verify that Eq.~\eqref{cons_of_magic} is the necessary condition for \(\bra{ij}[\mathcal{T},\rho_0]\ket{kl}=0+\mathcal{O}(\lambda^2)\); thus, conservation of magic is equivalent to the vanishing of the commutator \([\mathcal{T},\rho_0]\). This is in contrast to other information-theoretic measures, such as the fidelity or von Neumann entropy, which are trivially conserved at leading order, and thus do not lead to an interesting constraint. On the other hand, the vanishing of this commutator will surely entail the leading-order conservation of various other information-theoretic functions; thus, if one were attempting to formulate a generalised quantum-information principle applicable to a host of measures, one might be naturally inclined to this scenario. We hereafter treat the commutativity of \(\rho_0\) and \(\mathcal{T}\) as our guiding constraint.

In particular, Eq.~\eqref{cons_of_magic} is a special case of the conditions previously derived for minimal entanglement (for the case of an initial product state in flavour space), which we display here for completeness~\cite{Carena:2023vjc,Chang:2024wrx,McGinnis:2025brt}:
\begin{subequations}
    \label{min_ent}
    \begin{align}
        &M_{ijkl}=0\text{ for }i\neq k\wedge j\neq l \ ; \\
        &M_{ijil}=M_{kjkl}\text{ for }j\neq l \ ; \\
        &M_{ijlj}=M_{iklk}\text{ for }i\neq l \ ; \\
        &M_{ijij}+M_{klkl}=M_{ilil}+M_{jkjk} \ .
    \end{align}
\end{subequations}
From an information-theoretic perspective, the above conditions relate the \(\mathcal{S}\)-matrix to the Identity gate up to single-qubit operations, whereas Eq.~\eqref{cons_of_magic} forces the \(\mathcal{S}\)-matrix to \textit{be} the Identity gate. This may be seen more clearly if we display the \(\mathcal{O}(\lambda)\) scattering amplitudes in matrix form like so, following Ref.~\cite{Chang:2024wrx}:
\begin{equation}
    \label{display_ms}
    iM^0\equiv i\begin{pmatrix}
        M_{0000} & M_{0001} & M_{0010} & M_{0011} \\
        M_{0100} & M_{0101} & M_{0110} & M_{0111} \\
        M_{1000} & M_{1001} & M_{1010} & M_{1011} \\
        M_{1100} & M_{1101} & M_{1110} & M_{1111}
    \end{pmatrix}.
\end{equation}
Thus, the condition of Eq.~\eqref{cons_of_magic} amounts to requiring the above matrix be proportional to the identity matrix.

For \(\Phi^+\Phi^0\to\Phi^+\Phi^0\) scattering in the 2HDM, we have
\begin{equation}
    \label{0p->0p_2HDM}
    iM^0=i\begin{pmatrix}
       2\lambda_1 & \lambda_6 & \lambda_6 & \lambda_5 \\
       \lambda_6 & \lambda_3 & \lambda_4 & \lambda_7 \\
       \lambda_6 & \lambda_4 & \lambda_3 & \lambda_7 \\
       \lambda_5 & \lambda_7 & \lambda_7 & 2\lambda_2
    \end{pmatrix}.
\end{equation}
Eq.~\eqref{cons_of_magic} requires \(\lambda_1=\lambda_2=\lambda_3/2\equiv\lambda\) and \(\lambda_{i}=0\) otherwise, yielding the maximally symmetric quartic potential
\begin{equation}
    \label{p0top0pot2H}
    V_4=\lambda\left(\Phi_1^\dagger\Phi_1+\Phi_2^\dagger\Phi_2\right)^2.
\end{equation}
In this case, our potential clearly satisfies the conditions for natural alignment, where the left-hand side of Eq.~\eqref{align_cond_gen} vanishes identically for arbitrary \(t_\beta\)~\cite{Darvishi:2023fjh}. As the quartic potential is maximally symmetric, we note that we do not have spontaneous CP violation; we are thus justified to apply the conditions of Eq.~\eqref{align_cond_gen}. In our framework, the mass parameters of the Higgs potential do not contribute to the scattering amplitudes of Eq.~\eqref{0p->0p_2HDM}, leaving the quadratic sector unconstrained. However, having absorbed the mixing term \(m_{12}^2\) via an \(\text{SU}(2)\) rotation as described above, the full potential exhibits an accidental \(\mathrm{SO}(4)_1 \otimes \mathrm{SO}(4)_2\) symmetry reflecting invariance under independent rotations mixing the real degrees of freedom within each doublet \(\Phi_i\). This group acts to stabilise the lightest state from each of \(\Phi_1\) and \(\Phi_2\), raising the question of whether a symmetry of the kind required to stabilise dark matter could arise from information-theoretic considerations. An exploration of this possibility would require our toy model to be properly embedded in the SM, and as such, is beyond the scope of this work; see Ref.~\cite{Carena:2025wyh} for an investigation of mirror symmetry from maximisation of entanglement. If \(m_{12}^2\) is non-zero and cannot be rotated away, it breaks the symmetry to the diagonal subgroup \(\mathrm{SO}(4)_{1+2}\), acting identically on both doublets.

It is instructive to contrast this result with the conditions for minimal entanglement; in the 2HDM, and at the level of the \(\mathcal{O}(\lambda)\) contact diagram, Eq.~\eqref{min_ent} yields
\begin{equation}
    \lambda_1+\lambda_2=\lambda_3 \ ,\quad\lambda_4=\lambda_5=0 \ ,\quad\lambda_6=\lambda_7 \ .
\end{equation}
Comparing to Eq.~\eqref{align_cond_gen}, we see that these conditions are insufficient to realise natural alignment. Thus, Eq.~\eqref{cons_of_magic} is more constraining than minimisation of entanglement.

\section{Emergent symmetry from \texorpdfstring{\(\Phi \Phi \to \Phi \Phi\)}{Phi Phi to Phi Phi} scattering}
\label{sec:general}
We now extend our analysis from the specific process \(\Phi^+\Phi^0 \to \Phi^+\Phi^0\) to all \(2 \to 2\) scalar scattering channels involving states with total particle number \(\mathcal{P} = 2\), generically represented as \(\Phi_i \Phi_j \to \Phi_k \Phi_l\). We begin by extending the scattering formalism: we enlarge the Hilbert space to include a discrete isospin subspace to accommodate all scattering channels, and account for Bose symmetry in the initial state. We show that our commutator condition implies an \(\mathrm{SO}(8)\) symmetry acting on the quartic potential when scattering the most general allowed initial state, but this is reduced to an emergent \(\mathrm{SU}(2)_R\) symmetry when scattering two-particle states of definite isospin.

\subsection{Generalised Scattering Formalism}
\label{sec:setup}

It will be useful to rewrite Eq.~\eqref{2HDM_gen_potential} in terms of scalar bilinears that transform in a well-defined way under permutations of both Higgs-flavour and weak-isospin indices, so as to facilitate the construction of Bose-symmetric states. The four possible bilinears written in this way will be constructed from the singlet and triplet representations of \(\mathrm{SU}(2)\): the regular \(\mathrm{SU}(2)_L\) in the case of weak isospin, and a group we label \(\mathrm{SU}(2)_R\) only pre-emptively, corresponding to the freedom to change the Higgs-flavour basis. We can thus construct a complex bidoublet \(\Phi \equiv (\Phi_1, \Phi_2)\), which furnishes the \((\mathbf{2},\mathbf{2})\) product irrep under \(\mathrm{SU}(2)_L\otimes\mathrm{SU}(2)_R\). The \(\Phi\Phi\) bilinear then transforms as
\begin{equation}
\label{eq:decomposition}
\begin{aligned}
(\mathbf{2}_L\otimes\mathbf{2}_R)\otimes(\mathbf{2}_L\otimes\mathbf{2}_R)
&=(\mathbf{2}\otimes\mathbf{2})_L\otimes(\mathbf{2}\otimes\mathbf{2})_R \\
&= (\mathbf{1} \oplus \mathbf{3})_L \otimes (\mathbf{1} \oplus \mathbf{3})_R \\
&= (\mathbf{1}_L \otimes \mathbf{1}_R) \oplus (\mathbf{3}_L \otimes \mathbf{3}_R) \ .
\end{aligned}
\end{equation}
Note that the cross terms \((\mathbf{1}_L \otimes \mathbf{3}_R)\) and \((\mathbf{3}_L \otimes \mathbf{1}_R)\) vanish identically, since they are antisymmetric under the interchange of two identical scalar multiplets. 

In terms of these bilinears, the quartic part of the potential \(V_4\) can be written
\begin{equation}
  \label{eq:2hdm-potential-us}
   V_4 = \sum_{I=0}^{1} \sum_{m=-I}^{I} \sum_{R=0}^{1} \sum_{n, n^\prime = -R}^R  \lambda^{R}_{n,n^\prime} (\mathcal{O}_{m,n}^{RI})^\dagger \mathcal{O}_{m,n^\prime}^{RI}
\end{equation}
where
\begin{equation} \label{eq:operator-2}
  \mathcal{O}_{m,n}^{RI} =\sum_{a,b} \sum_{i,j} C^{I,m}_{ab} C^{R,n}_{ij} \Phi_{ia} \Phi_{jb} \ .
\end{equation}
Here, the \(C^{I,m}_{ab}\) and \( C^{R,n}_{ij}\) are Clebsch--Gordan coefficients coupling the bilinear \(\mathbf{2} \otimes \mathbf{2}\) representation to generalised-spin-\(I\) and generalised-spin-\(R\) irreps with projections \(m\) and \(n\). The indices \(i,j\) enumerate flavour and \(a,b\) are \(\mathrm{SU}(2)_L\) fundamental indices. The matrices are
\begin{align}
\mathbf{C}^{0,0} &= 
\frac{1}{\sqrt{2}}
\begin{pmatrix}
  0 & 1 \\
  -1 & 0
\end{pmatrix} \ , \\
\mathbf{C}^{1,m} &= 
\begin{cases}
\begin{pmatrix}
  1 & 0 \\
  0 & 0
\end{pmatrix} & m=1 \\
\frac{1}{\sqrt{2}}
\begin{pmatrix}
  0 & 1 \\
  1 & 0
\end{pmatrix} & m = 0 \\
\begin{pmatrix}
  0 & 0 \\
  0 & 1
\end{pmatrix} & m=-1
\end{cases} \ .
\end{align}
The coefficients \(\lambda_{n,n^\prime}^R\) contain ten real parameters in total: one in \(\lambda^0_{0,0} \equiv \lambda_S\) and nine within the \(3\times 3\) Hermitian matrix \(\boldsymbol{\lambda}^1 \equiv \boldsymbol{\lambda}_T\). We emphasise that this amounts at this stage only to a suggestive change of basis on the potential. Eq.~\eqref{eq:2hdm-potential-us} is not invariant under \(\mathrm{SU}(2)_R\) since \(\lambda^R_{n,n^\prime}\) carries free \(\mathrm{SU}(2)_R\) indices and in general is not proportional to the identity. The correspondence between the parameters in this potential and those of Eq.~\eqref{2HDM_gen_potential} is
\begin{equation}
 \begin{aligned}
  \label{eq:couplings}
  \lambda_S &= \tfrac{1}{2}(\lambda_3-\lambda_4) \\ 
  \boldsymbol{\lambda}_T &= 
  \begin{pmatrix} 
   \lambda_1 & \tfrac{1}{\sqrt{2}}\lambda_6 & \tfrac{1}{2}\lambda_5 \\
   \tfrac{1}{\sqrt{2}}\lambda_6^* & \tfrac{1}{2} (\lambda_3+\lambda_4) & \tfrac{1}{\sqrt{2}}\lambda_7 \\
   \tfrac{1}{2}\lambda_5^* & \tfrac{1}{\sqrt{2}}\lambda_7^* & \lambda_2
  \end{pmatrix} \ ,
 \end{aligned}
\end{equation}
where we order the triplet basis states \(|1,n\rangle\) in order of decreasing \(n\), i.e., as \(\{ \ket{1,1}, \ket{1,0},|1,-1 \rangle \}\).

We refine our partition of the Hilbert space to \(\mathcal{H}_2\simeq \mathcal{H}_\mathbf{p}\otimes \mathcal{H}_\text{iso} \otimes \mathcal{H}_\text{flav}\), where the new factor \(\mathcal{H}_\text{iso}\simeq \mathbb{C}^2\otimes\mathbb{C}^2\) corresponds to the discrete two-qubit isospin space. We indicate a generic basis for \(\mathcal{H}_2\) as \(|ab\rangle \otimes  \ket{ij} \otimes | \mathbf{p}_1 \mathbf{p}_2 \rangle\), where \(\mathbf{p}_1, \mathbf{p}_2\) are three-momenta associated with the momentum four-vectors \(p_{1,2} = (E_{1,2}, \mathbf{p}_{1,2})\), the indices \(a,b\) are fundamental \(\mathrm{SU}(2)_L\) isospin indices, and \(i,j\) are Higgs flavour indices. The momentum and discrete states are normalised in the standard way as above. A general initial product state in the isospin, flavour and momentum subspaces of \(\mathcal{H}_2\) now takes the form
\begin{equation} \label{eq:product-state}
\begin{aligned}
| \psi_0 \rangle &=  \sum_{a,b} A_{ab} |ab\rangle \otimes \sum_{i,j} B_{ij}   |ij\rangle\\
&\quad \otimes \int \frac{\mathrm{d}^3 \mathbf{p}_1}{(2\pi)^3 2E_1} \frac{\mathrm{d}^3 \mathbf{p}_2}{(2\pi)^3 2E_2}\phi(\mathbf{p}_1, \mathbf{p}_2) | \mathbf{p}_1 \mathbf{p}_2 \rangle \ ,
\end{aligned}
\end{equation}
where, again, the coefficients \(A_{ab}\) and \(B_{ij}\) and the momentum wave function \(\phi(\mathbf{p}_1,\mathbf{p}_2)\) satisfy the usual normalisation conditions.

The products \(| a\rangle \otimes | b\rangle = | ab \rangle\) and \(|i\rangle \otimes | j\rangle = | i j \rangle\) can be decomposed using Clebsch--Gordan coefficients, so that the general initial state may alternately be written
\begin{equation} \label{eq:product-state-cgcs}
\begin{aligned}
\ket{\psi_0} &= \sum_{I,m} \alpha_{I,m} \ket{I,m} \otimes \sum_{R,n} \beta_{R,n} \ket{R,n}\\
&\quad \otimes \int \frac{\mathrm{d}^3 \mathbf{p}_1}{(2\pi)^3 2E_1} \frac{\mathrm{d}^3 \mathbf{p}_2}{(2\pi)^3 2E_2}\phi(\mathbf{p}_1, \mathbf{p}_2) | \mathbf{p}_1 \mathbf{p}_2 \rangle \ .
\end{aligned}
\end{equation}
Here we have adopted the standard notation familiar from the study of angular momentum, where \(| J, m\rangle\) is used for irreps of \(\mathrm{SU}(2)\) labelled by their total weight \(J\) and third-component projection \(m\). The \(\alpha_{I, m}\) and \(\beta_{R, n}\) are coefficients related to the \(A_{ab}\) and \(B_{ij}\) from Eq.~\eqref{eq:product-state} respectively.

Now that we are treating weak isospin as an additional quantum number, the states involved in the scattering are of the same underlying species. Bose symmetry thus requires that \(\ket{\psi_0}\) be symmetric under interchange of the two particles. For now we assume \(\phi(\mathbf{p}_1,\mathbf{p}_2)\) is symmetric under \(\mathbf{p}_1 \leftrightarrow \mathbf{p}_2\) and suppress the momentum part of the state going forward for clarity. We return to the case of an antisymmetric momentum wave function in Sec.~\ref{sec:isospin}. This assumption enforces that only the products shown in Eq.~\eqref{eq:decomposition} can enter in Eq.~\eqref{eq:product-state-cgcs}, i.e.:
\begin{equation} \label{eq:in-state}
\begin{aligned}
    \ket{\psi_0} &= \alpha_{0,0} \ket{0,0} \otimes \beta_{0,0} \ket{0,0} \\
    &+ \alpha_{1,m} \ket{1,m} \otimes \beta_{1,n} \ket{1,n} \ ,
\end{aligned}
\end{equation}
with an implied sum over the projections \(m\) and \(n\). Bose symmetry has thus introduced a level of entanglement between the isospin and flavour subspaces. We can construct a general initial-state density matrix for the product state as
\begin{equation} \label{eq:rho-in}
\begin{aligned}
\rho &=
|\alpha_{0,0}|^2 \ketbra{0,0}{0,0} \otimes |\beta_{0,0}|^2 \ketbra{0,0}{0,0} \\
&+ \alpha_{0,0} \alpha_{1,m'}^* \ketbra{0,0}{1,m'} \otimes \beta_{0,0} \beta_{1,n'}^* \ketbra{0,0}{1,n'} \\
&+ \alpha_{1,m} \alpha_{0,0}^* \ketbra{1,m}{0,0} \otimes \beta_{1,n} \beta_{0,0}^* \ketbra{1,n}{0,0} \\
&+ \alpha_{1,m} \alpha_{1,m'}^* \ketbra{1,m}{1,m'} \otimes \beta_{1,n} \beta_{1,n'}^* \ketbra{1,n}{1,n'} \ ,
\end{aligned}
\end{equation}
where the operators are written in the order \(\mathrm{SU}(2)_L \otimes \mathrm{SU}(2)_R\).

Since the operator \(\mathcal{O}_{m,n}^{RI}\) of Eq.~\eqref{eq:operator-2} annihilates the two-particle state \(\ket{s}\equiv| I,m \rangle \otimes |R,n \rangle\), and the conjugate operator \((\mathcal{O}_{m,n}^{RI})^\dagger\) creates it, the discrete part of the \(\mathcal{T}\) matrix for \(\Phi\Phi \to \Phi\Phi\) scattering can be written in a form convenient for comparison with the density matrix:
\begin{align} 
  \mathcal{T} &=\sum_{s,s'}\ket{s}\braket{s|\mathcal{T}|s'}\bra{s'} =\sum_{s,s'}\mathcal{M}_{s\to s'}\ket{s}\bra{s'} \\
  &=\lambda_{S} \ketbra{0,0}{0,0}  \otimes \ketbra{0,0}{0,0} \nonumber\\ & \ + \sum_{m,n,n^\prime} (\lambda_{T})_{n,n^\prime} \ketbra{1,m}{1,m} \otimes \ketbra{ 1, n}{1,n^\prime} \ , \label{eq:t-matrix}
\end{align}
reading off the tree-level amplitudes as the Lagrangian parameters \(\lambda_S\) and \(\boldsymbol{\lambda}_T\) defined in Eq.~\eqref{eq:couplings}.

\subsection{Maximal symmetry from \texorpdfstring{\(\Phi \Phi \to \Phi \Phi\)}{Phi Phi to Phi Phi} scattering}
\label{sec:phiplus-phi0-scattering}

We now consider the scattering of the generic initial product state described by Eq.~\eqref{eq:rho-in} in full generality. The commutator can be written in \(2 \times 2\) block form; in the basis \(\{\ket{0,0}\otimes\ket{0,0}, \ket{1,m}\otimes\ket{1,n}\}\) we have
\begin{equation} \label{eq:commutator}
\begin{aligned}
\Bigg[
&\begin{pmatrix}
\lambda_{S} & 0\\
0 & (\lambda_T)_{n,n'}\delta_{m,m'} \\
\end{pmatrix}, \\ 
&\begin{pmatrix}
|\alpha_{0,0}|^2 |\beta_{0,0}|^2 & \alpha_{0,0}\alpha_{1,m'}^*\beta_{0,0} \beta_{1,n'}^* \\
\alpha_{1,m}\alpha_{0,0}^*\beta_{1,n} \beta_{0,0}^* & \alpha_{1,m}\alpha_{1,m'}^*\beta_{1,n} \beta_{1,n'}^*
\end{pmatrix}
\Bigg] \ .
\end{aligned}
\end{equation}
This vanishes for arbitrary \(\alpha_{0,0},\alpha_{1,n},\beta_{0,0},\beta_{1,n}\) if and only if the upper-left and lower-right blocks of \(\mathcal{T}\) act proportionally to the identity on the subspace, i.e., when \(\lambda_S = (\lambda_T)_{n,n}\) with no sum over \(n\). This fixes \(\lambda_1 = \lambda_2 = \tfrac{1}{2}\lambda_3\) from Eq.~\eqref{eq:couplings}, resulting in the emergence of the maximal \(\mathrm{SO}(8)\) symmetry on \(V_4\).

Armed with this general result, we revisit the case of \(\Phi^+\Phi^0\) scattering studied in Sec.~\ref{magic_cons_align} in our isospin-covariant formalism. The \(\mathrm{SU}(2)_L\) triplet and singlet combinations for \(\mathcal{P}=2\) scattering, written in terms of the charged states, are
\begin{align}
\ket{1,m} &\sim 
\begin{pmatrix}
 \ket{\Phi^+ \Phi^+} \\
  \frac{1}{\sqrt{2}} (\ket{\Phi^+ \Phi^0} + \ket{\Phi^0 \Phi^+}) \\
  \ket{\Phi^0 \Phi^0} \\
\end{pmatrix}_m\\
\ket{0,0}  &\sim  \frac{1}{\sqrt{2}} (\ket{\Phi^+ \Phi^0} - \ket{\Phi^0 \Phi^+})
\end{align}
and flavour indices are left implied. The \(\mathrm{SU}(2)_L\) part of the state \(\ket{\Phi^+ \Phi^0}\) can thus be written
\begin{equation} \label{eq:phiPphi0-decomp}
\ket{\Phi^+ \Phi^0} = \frac{1}{\sqrt{2}} ( \ket{1,0} + \ket{0,0} )
\end{equation}
for which  \(\alpha_{0,0} = \alpha_{1,0} = 1/\sqrt{2}\). In this particular case, the commutator reduces to
\begin{equation}
\begin{aligned}
    \frac{1}{2}\Bigg[
    &\begin{pmatrix}
    \lambda_{S} & 0\\
    0 & (\lambda_T)_{n,n'}\delta_{m,m'} \\
    \end{pmatrix}, \\
    &\begin{pmatrix}
    |\beta_{0,0}|^2 & \delta_{0,m'}\beta_{0,0} \beta_{1,n'}^* \\
    \delta_{0,m}\beta_{1,n} \beta_{0,0}^* & \delta_{0,m}\delta_{0,m'}\beta_{1,n} \beta_{1,n'}^*
    \end{pmatrix}\Bigg]
\end{aligned}
\end{equation}
but for arbitrary \(\beta_{0,0},\beta_{1,n}\), it remains true that the only circumstance in which this object can vanish is that described above, recovering the maximal symmetry. 

Essential to this result is the fact that the initial state is a superposition of states of differing total isospin, as seen clearly in Eq.~\eqref{eq:phiPphi0-decomp}. In the next section, we go on to show that scattering states of \emph{definite} isospin results in a smaller \(\mathrm{SU}(2)_R\) symmetry due to the entanglement between the two discrete sectors imposed by Bose symmetry.

\subsection{Recovering \texorpdfstring{\(\mathrm{SU}(2)_R\)}{SU2R}}
\label{sec:isospin}

We now turn to the general case of \(\mathcal{P}=2\) scattering of states of definite isospin. We begin with the triplet state, whose initial density matrix is given by Eq.~\eqref{eq:rho-in} with \(\alpha_{00}=\beta_{00}=0\):
\begin{equation}
\rho_0 = \alpha_{1,m}\alpha_{1,m'}^* \ketbra{1,m}{1,m'} \otimes \beta_{1,n} \beta_{1,n'}^* \ketbra{1,n}{1,n'}
\end{equation}
It is clear that \(\rho_0\) trivially commutes with the singlet piece of \(\mathcal{T}\) as seen in Eq.~\eqref{eq:t-matrix}. Thus, in this case the commutation condition enforces \(\boldsymbol{\lambda}_T \sim \mathbbm{1}_{3}\) for arbitrary \(\alpha_{1,m}\) and \(\beta_{1,n}\). We see in Eq.~\eqref{eq:couplings} that this imposes the relations
\begin{equation}
\lambda_1 = \lambda_2 = \tfrac{1}{2} (\lambda_3 + \lambda_4)
\end{equation}
on the quartic part of the potential. In the presence of these relations, \(V_4\) becomes invariant under a global \(\mathrm{SU}(2)\) rotation acting in the Higgs-flavour space~\cite{Darvishi:2019dbh}. This is analogous to the quartic potential of the scalar bidoublet present in models featuring left--right symmetry. Importantly, the full transition matrix carries a non-trivial structure, as only the triplet block is proportional to the identity; we also note that the alignment limit is still realised, as per Eq.~\eqref{align_cond_gen}.

The case of singlet scattering produces no additional relations. Setting \(\alpha_{1,m}=\beta_{1,n}=0\), the density matrix is
\begin{equation}
\rho_0 = |\alpha_{0,0}|^2 \ketbra{0,0}{0,0} \otimes |\beta_{0,0}|^2 \ketbra{0,0}{0,0} \ ,
\end{equation}
and this trivially commutes with \(\mathcal{T}\) for all \(\alpha_{00}\) and \(\beta_{00}\). 

The recovery of an \(\mathrm{SU}(2)_R\) flavour symmetry here is perhaps unsurprising. Already in Eq.~\eqref{eq:in-state} we see that Bose symmetry has enforced a correlation between the \(\mathrm{SU}(2)\) representations in each term. Choosing to scatter an isospin triplet, for instance, automatically selects the spin-1 irrep on the flavour side as well. Requiring the \(\mathcal{T}\)-matrix to commute with irreps of \(\mathrm{SU}(2)_R\) then imprints the symmetry onto the quartic potential. 

\subsection{Antisymmetric momentum wave function}

Until now we have assumed that the momentum wave function of Eq.~\eqref{eq:product-state} is symmetric under \(\mathbf{p}_1 \leftrightarrow \mathbf{p}_2\). We now revisit the antisymmetric case to tie off the last loose end. In the case of antisymmetric \(\phi(\mathbf{p}_1, \mathbf{p}_2)\), the remainder of the state should be antisymmetric under particle interchange. This singles out the complementary parts of the decomposition of Eq.~\eqref{eq:decomposition} that are totally antisymmetric. The initial antisymmetric product state is then
\begin{equation}
| \psi_{0,A} \rangle = \alpha_{0,0} \ket{0,0} \otimes \beta_{1,n} \ket{1,n} + \alpha_{1,m} \ket{1,m} \otimes \beta_{0,0} \ket{0,0} \ ,
\end{equation}
with
\begin{equation}
\begin{aligned} \label{eq:antisym-rho}
\rho_{0,A} &=
|\alpha_{0,0}|^2 \ketbra{0,0}{0,0} \otimes \beta_{1,n} \beta_{1,n'}^* \ketbra{1,n}{1,n'} \\
&+ \alpha_{0,0} \alpha_{1,m'}^* \ketbra{0,0}{1,m'} \otimes \beta_{1,n} \beta_{0,0}^* \ketbra{1,n}{0,0} \\
&+ \alpha_{1,m} \alpha_{0,0}^* \ketbra{1,m}{0,0} \otimes \beta_{0,0} \beta_{1,n'}^* \ketbra{0,0}{1,n'} \\
&+ \alpha_{1,m}\alpha_{1,m'}^* \ketbra{1,m}{1,m'} \otimes |\beta_{0,0}|^2 \ketbra{0,0}{0,0} \ .
\end{aligned}
\end{equation}
Comparing to Eq.~\eqref{eq:t-matrix}, it is clear that \(\mathcal{T}\) and \(\rho_{0,A}\) are orthogonal for all choices of \(\alpha_{I,m}\) and \(\beta_{R,n}\): any non-zero amplitude in the isospin subspace is paired with a vanishing one in the flavour subspace, and vice versa. We therefore conclude that scattering with an antisymmetric momentum wave function imposes no further constraints on the quartic couplings.

\section{Conclusions}
\label{conclusion}
In this work we have explored quantum information-theoretic constraints in the \(2\to 2\) scattering of scalars in a two-Higgs-doublet model. Building on previous studies that considered entanglement extremisation, we have introduced a new perspective based on conservation of magic or, equivalently, commutativity of the \(\mathcal{T}\)-matrix with the in-state density matrix as a guiding principle for emergent symmetries and field-theoretic structure.

We demonstrated that imposing magic conservation at lowest order in perturbation theory for elastic $\Phi^+ \Phi^0\to \Phi^+\Phi^0$ scattering constrains the quartic part of the Higgs potential to a maximally symmetric form. This would naturally realise the alignment limit in a full 2HDM extension of the Standard Model. We found that the constraints on the scattering amplitude imposed by our principle are a special case of those previously derived in the literature from the minimisation of entanglement for initial product states. Our conclusions suggest this commutativity requirement be viewed as a useful explanatory tool, which here appears to provide a quantum information-theoretic justification for an otherwise unexplained fine-tuning.

We then extended our analysis to all \(2\to 2\) scalar scattering processes with total particle number \(\mathcal{P}=2\) by treating (the third component of) weak isospin as an explicit qubit degree of freedom in its own right. In this case, Bose symmetry constrains the form of the initial-state density matrix in the finite part of the Hilbert space; we have demonstrated that the maximal symmetry is recovered when scattering particular or general superpositions of definite-isospin states, but that when scattering states of definite isospin, we recover a smaller \(\mathrm{SU}(2)_R\) symmetry that nevertheless admits Standard Model alignment. These conclusions hold regardless of the symmetry properties of the initial-state momentum wave function.

We comment here on the consistency of our results with previous studies. Even though magic conservation is more constraining than entanglement minimisation, the results of Ref.~\cite{Chang:2024wrx} are not inconsistent with our results because we do not consider the same class of scattering processes; namely, we concentrate only on $\mathcal{P}=2$ channels. We anticipate that the inclusion of isospin as an additional quantum number may lead to different results even in the $\mathcal{P}=0$ channels too. Ref.~\cite{Kowalska:2024kbs} studies other $\mathcal{P}=2$ channels in the context of entanglement minimisation, but their results aggregate over all scattering processes; further, we find that symmetrisation of the initial state (for the case of identical particles) plays an important role in our analysis.

Future work should extend this analysis beyond lowest order in perturbation theory. This requires the inclusion of multi-particle states and would necessitate a more careful treatment of the scattering formalism used here. A complete treatment of the total particle number \(\mathcal{P}=0\) scalar scattering processes omitted here would also be illuminating. Interactions with gauge bosons and fermions should be incorporated in order to understand the interplay of this framework with the Standard Model --- in particular, whether phenomenologically interesting scenarios (e.g., an inert doublet model, a canonical left--right symmetric model, etc.) can be realised. In all, it is clear that candidate quantum information-theoretic principles extending beyond entanglement provide useful constraints on field-theoretic structure; we believe that such ideas are worth exploring in more detail going forward.

\section*{Acknowledgements}
We thank Rafael Aoude, Hannah Banks, Tomas L.\ Howson, Raymond R.\ Volkas and Chris D.\ White for useful conversations and comments on the draft. All authors are supported by the Australian Research Council grant CE200100008. MJW is further supported by the Australian Research Council grant DP220100007. ENVW acknowledges the support he has received for this research through the provision of an Australian Government Research Training Program Scholarship.

\bibliographystyle{apsrev4-1}
\bibliography{main}

\begin{thebibliography}{28}%
\makeatletter
\providecommand \@ifxundefined [1]{%
 \@ifx{#1\undefined}
}%
\providecommand \@ifnum [1]{%
 \ifnum #1\expandafter \@firstoftwo
 \else \expandafter \@secondoftwo
 \fi
}%
\providecommand \@ifx [1]{%
 \ifx #1\expandafter \@firstoftwo
 \else \expandafter \@secondoftwo
 \fi
}%
\providecommand \natexlab [1]{#1}%
\providecommand \enquote  [1]{``#1''}%
\providecommand \bibnamefont  [1]{#1}%
\providecommand \bibfnamefont [1]{#1}%
\providecommand \citenamefont [1]{#1}%
\providecommand \href@noop [0]{\@secondoftwo}%
\providecommand \href [0]{\begingroup \@sanitize@url \@href}%
\providecommand \@href[1]{\@@startlink{#1}\@@href}%
\providecommand \@@href[1]{\endgroup#1\@@endlink}%
\providecommand \@sanitize@url [0]{\catcode `\\12\catcode `\$12\catcode
  `\&12\catcode `\#12\catcode `\^12\catcode `\_12\catcode `\%12\relax}%
\providecommand \@@startlink[1]{}%
\providecommand \@@endlink[0]{}%
\providecommand \url  [0]{\begingroup\@sanitize@url \@url }%
\providecommand \@url [1]{\endgroup\@href {#1}{\urlprefix }}%
\providecommand \urlprefix  [0]{URL }%
\providecommand \Eprint [0]{\href }%
\providecommand \doibase [0]{http://dx.doi.org/}%
\providecommand \selectlanguage [0]{\@gobble}%
\providecommand \bibinfo  [0]{\@secondoftwo}%
\providecommand \bibfield  [0]{\@secondoftwo}%
\providecommand \translation [1]{[#1]}%
\providecommand \BibitemOpen [0]{}%
\providecommand \bibitemStop [0]{}%
\providecommand \bibitemNoStop [0]{.\EOS\space}%
\providecommand \EOS [0]{\spacefactor3000\relax}%
\providecommand \BibitemShut  [1]{\csname bibitem#1\endcsname}%
\let\auto@bib@innerbib\@empty
\bibitem [{\citenamefont {McGinnis}(2025)}]{McGinnis:2025brt}%
  \BibitemOpen
  \bibfield  {author} {\bibinfo {author} {\bibfnamefont {N.}~\bibnamefont
  {McGinnis}},\ }\href@noop {} {\  (\bibinfo {year} {2025})},\ \Eprint
  {http://arxiv.org/abs/2504.21079} {arXiv:2504.21079 [hep-th]} \BibitemShut
  {NoStop}%
\bibitem [{\citenamefont {Beane}\ \emph {et~al.}(2019)\citenamefont {Beane},
  \citenamefont {Kaplan}, \citenamefont {Klco},\ and\ \citenamefont
  {Savage}}]{Beane:2018oxh}%
  \BibitemOpen
  \bibfield  {author} {\bibinfo {author} {\bibfnamefont {S.~R.}\ \bibnamefont
  {Beane}}, \bibinfo {author} {\bibfnamefont {D.~B.}\ \bibnamefont {Kaplan}},
  \bibinfo {author} {\bibfnamefont {N.}~\bibnamefont {Klco}}, \ and\ \bibinfo
  {author} {\bibfnamefont {M.~J.}\ \bibnamefont {Savage}},\ }\href {\doibase
  10.1103/PhysRevLett.122.102001} {\bibfield  {journal} {\bibinfo  {journal}
  {Phys. Rev. Lett.}\ }\textbf {\bibinfo {volume} {122}},\ \bibinfo {pages}
  {102001} (\bibinfo {year} {2019})},\ \Eprint
  {http://arxiv.org/abs/1812.03138} {arXiv:1812.03138 [nucl-th]} \BibitemShut
  {NoStop}%
\bibitem [{\citenamefont {Low}\ and\ \citenamefont
  {Mehen}(2021)}]{Low:2021ufv}%
  \BibitemOpen
  \bibfield  {author} {\bibinfo {author} {\bibfnamefont {I.}~\bibnamefont
  {Low}}\ and\ \bibinfo {author} {\bibfnamefont {T.}~\bibnamefont {Mehen}},\
  }\href {\doibase 10.1103/PhysRevD.104.074014} {\bibfield  {journal} {\bibinfo
   {journal} {Phys. Rev. D}\ }\textbf {\bibinfo {volume} {104}},\ \bibinfo
  {pages} {074014} (\bibinfo {year} {2021})},\ \Eprint
  {http://arxiv.org/abs/2104.10835} {arXiv:2104.10835 [hep-th]} \BibitemShut
  {NoStop}%
\bibitem [{\citenamefont {Liu}\ \emph {et~al.}(2023)\citenamefont {Liu},
  \citenamefont {Low},\ and\ \citenamefont {Mehen}}]{Liu:2022grf}%
  \BibitemOpen
  \bibfield  {author} {\bibinfo {author} {\bibfnamefont {Q.}~\bibnamefont
  {Liu}}, \bibinfo {author} {\bibfnamefont {I.}~\bibnamefont {Low}}, \ and\
  \bibinfo {author} {\bibfnamefont {T.}~\bibnamefont {Mehen}},\ }\href
  {\doibase 10.1103/PhysRevC.107.025204} {\bibfield  {journal} {\bibinfo
  {journal} {Phys. Rev. C}\ }\textbf {\bibinfo {volume} {107}},\ \bibinfo
  {pages} {025204} (\bibinfo {year} {2023})},\ \Eprint
  {http://arxiv.org/abs/2210.12085} {arXiv:2210.12085 [quant-ph]} \BibitemShut
  {NoStop}%
\bibitem [{\citenamefont {Liu}\ and\ \citenamefont {Low}(2024)}]{Liu:2023bnr}%
  \BibitemOpen
  \bibfield  {author} {\bibinfo {author} {\bibfnamefont {Q.}~\bibnamefont
  {Liu}}\ and\ \bibinfo {author} {\bibfnamefont {I.}~\bibnamefont {Low}},\
  }\href {\doibase 10.1016/j.physletb.2024.138899} {\bibfield  {journal}
  {\bibinfo  {journal} {Phys. Lett. B}\ }\textbf {\bibinfo {volume} {856}},\
  \bibinfo {pages} {138899} (\bibinfo {year} {2024})},\ \Eprint
  {http://arxiv.org/abs/2312.02289} {arXiv:2312.02289 [hep-ph]} \BibitemShut
  {NoStop}%
\bibitem [{\citenamefont {Thaler}\ and\ \citenamefont
  {Trifinopoulos}(2025)}]{Thaler:2024anb}%
  \BibitemOpen
  \bibfield  {author} {\bibinfo {author} {\bibfnamefont {J.}~\bibnamefont
  {Thaler}}\ and\ \bibinfo {author} {\bibfnamefont {S.}~\bibnamefont
  {Trifinopoulos}},\ }\href {\doibase 10.1103/PhysRevD.111.056021} {\bibfield
  {journal} {\bibinfo  {journal} {Phys. Rev. D}\ }\textbf {\bibinfo {volume}
  {111}},\ \bibinfo {pages} {056021} (\bibinfo {year} {2025})},\ \Eprint
  {http://arxiv.org/abs/2410.23343} {arXiv:2410.23343 [hep-ph]} \BibitemShut
  {NoStop}%
\bibitem [{\citenamefont {Carena}\ \emph {et~al.}(2024)\citenamefont {Carena},
  \citenamefont {Low}, \citenamefont {Wagner},\ and\ \citenamefont
  {Xiao}}]{Carena:2023vjc}%
  \BibitemOpen
  \bibfield  {author} {\bibinfo {author} {\bibfnamefont {M.}~\bibnamefont
  {Carena}}, \bibinfo {author} {\bibfnamefont {I.}~\bibnamefont {Low}},
  \bibinfo {author} {\bibfnamefont {C.~E.~M.}\ \bibnamefont {Wagner}}, \ and\
  \bibinfo {author} {\bibfnamefont {M.-L.}\ \bibnamefont {Xiao}},\ }\href
  {\doibase 10.1103/PhysRevD.109.L051901} {\bibfield  {journal} {\bibinfo
  {journal} {Phys. Rev. D}\ }\textbf {\bibinfo {volume} {109}},\ \bibinfo
  {pages} {L051901} (\bibinfo {year} {2024})},\ \Eprint
  {http://arxiv.org/abs/2307.08112} {arXiv:2307.08112 [hep-ph]} \BibitemShut
  {NoStop}%
\bibitem [{\citenamefont {Chang}\ and\ \citenamefont
  {Jacobo}(2024)}]{Chang:2024wrx}%
  \BibitemOpen
  \bibfield  {author} {\bibinfo {author} {\bibfnamefont {S.}~\bibnamefont
  {Chang}}\ and\ \bibinfo {author} {\bibfnamefont {G.}~\bibnamefont {Jacobo}},\
  }\href {\doibase 10.1103/PhysRevD.110.096020} {\bibfield  {journal} {\bibinfo
   {journal} {Phys. Rev. D}\ }\textbf {\bibinfo {volume} {110}},\ \bibinfo
  {pages} {096020} (\bibinfo {year} {2024})},\ \Eprint
  {http://arxiv.org/abs/2409.13030} {arXiv:2409.13030 [hep-ph]} \BibitemShut
  {NoStop}%
\bibitem [{\citenamefont {Kowalska}\ and\ \citenamefont
  {Sessolo}(2024)}]{Kowalska:2024kbs}%
  \BibitemOpen
  \bibfield  {author} {\bibinfo {author} {\bibfnamefont {K.}~\bibnamefont
  {Kowalska}}\ and\ \bibinfo {author} {\bibfnamefont {E.~M.}\ \bibnamefont
  {Sessolo}},\ }\href {\doibase 10.1007/JHEP07(2024)156} {\bibfield  {journal}
  {\bibinfo  {journal} {JHEP}\ }\textbf {\bibinfo {volume} {07}},\ \bibinfo
  {pages} {156} (\bibinfo {year} {2024})},\ \Eprint
  {http://arxiv.org/abs/2404.13743} {arXiv:2404.13743 [hep-ph]} \BibitemShut
  {NoStop}%
\bibitem [{\citenamefont {Carena}\ \emph {et~al.}(2025)\citenamefont {Carena},
  \citenamefont {Coloretti}, \citenamefont {Liu}, \citenamefont {Littmann},
  \citenamefont {Low},\ and\ \citenamefont {Wagner}}]{Carena:2025wyh}%
  \BibitemOpen
  \bibfield  {author} {\bibinfo {author} {\bibfnamefont {M.}~\bibnamefont
  {Carena}}, \bibinfo {author} {\bibfnamefont {G.}~\bibnamefont {Coloretti}},
  \bibinfo {author} {\bibfnamefont {W.}~\bibnamefont {Liu}}, \bibinfo {author}
  {\bibfnamefont {M.}~\bibnamefont {Littmann}}, \bibinfo {author}
  {\bibfnamefont {I.}~\bibnamefont {Low}}, \ and\ \bibinfo {author}
  {\bibfnamefont {C.~E.~M.}\ \bibnamefont {Wagner}},\ }\href@noop {} {\
  (\bibinfo {year} {2025})},\ \Eprint {http://arxiv.org/abs/2505.00873}
  {arXiv:2505.00873 [hep-ph]} \BibitemShut {NoStop}%
\bibitem [{\citenamefont {Cervera-Lierta}\ \emph {et~al.}(2017)\citenamefont
  {Cervera-Lierta}, \citenamefont {Latorre}, \citenamefont {Rojo},\ and\
  \citenamefont {Rottoli}}]{Cervera-Lierta:2017tdt}%
  \BibitemOpen
  \bibfield  {author} {\bibinfo {author} {\bibfnamefont {A.}~\bibnamefont
  {Cervera-Lierta}}, \bibinfo {author} {\bibfnamefont {J.~I.}\ \bibnamefont
  {Latorre}}, \bibinfo {author} {\bibfnamefont {J.}~\bibnamefont {Rojo}}, \
  and\ \bibinfo {author} {\bibfnamefont {L.}~\bibnamefont {Rottoli}},\ }\href
  {\doibase 10.21468/SciPostPhys.3.5.036} {\bibfield  {journal} {\bibinfo
  {journal} {SciPost Phys.}\ }\textbf {\bibinfo {volume} {3}},\ \bibinfo
  {pages} {036} (\bibinfo {year} {2017})},\ \Eprint
  {http://arxiv.org/abs/1703.02989} {arXiv:1703.02989 [hep-th]} \BibitemShut
  {NoStop}%
\bibitem [{\citenamefont {N\'u\~nez}\ \emph {et~al.}(2025)\citenamefont
  {N\'u\~nez}, \citenamefont {Cervera-Lierta},\ and\ \citenamefont
  {Latorre}}]{Nunez:2025dch}%
  \BibitemOpen
  \bibfield  {author} {\bibinfo {author} {\bibfnamefont {C.}~\bibnamefont
  {N\'u\~nez}}, \bibinfo {author} {\bibfnamefont {A.}~\bibnamefont
  {Cervera-Lierta}}, \ and\ \bibinfo {author} {\bibfnamefont {J.~I.}\
  \bibnamefont {Latorre}},\ }\href@noop {} {\  (\bibinfo {year} {2025})},\
  \Eprint {http://arxiv.org/abs/2504.15353} {arXiv:2504.15353 [hep-th]}
  \BibitemShut {NoStop}%
\bibitem [{\citenamefont {Zanardi}\ \emph {et~al.}(2000)\citenamefont
  {Zanardi}, \citenamefont {Zalka},\ and\ \citenamefont
  {Faoro}}]{Zanardi:2000zz}%
  \BibitemOpen
  \bibfield  {author} {\bibinfo {author} {\bibfnamefont {P.}~\bibnamefont
  {Zanardi}}, \bibinfo {author} {\bibfnamefont {C.}~\bibnamefont {Zalka}}, \
  and\ \bibinfo {author} {\bibfnamefont {L.}~\bibnamefont {Faoro}},\ }\href
  {\doibase 10.1103/PhysRevA.62.030301} {\bibfield  {journal} {\bibinfo
  {journal} {Phys. Rev. A}\ }\textbf {\bibinfo {volume} {62}},\ \bibinfo
  {pages} {030301} (\bibinfo {year} {2000})},\ \Eprint
  {http://arxiv.org/abs/quant-ph/0005031} {arXiv:quant-ph/0005031} \BibitemShut
  {NoStop}%
\bibitem [{\citenamefont {White}\ and\ \citenamefont
  {White}(2024)}]{White:2024nuc}%
  \BibitemOpen
  \bibfield  {author} {\bibinfo {author} {\bibfnamefont {C.~D.}\ \bibnamefont
  {White}}\ and\ \bibinfo {author} {\bibfnamefont {M.~J.}\ \bibnamefont
  {White}},\ }\href {\doibase 10.1103/PhysRevD.110.116016} {\bibfield
  {journal} {\bibinfo  {journal} {Phys. Rev. D}\ }\textbf {\bibinfo {volume}
  {110}},\ \bibinfo {pages} {116016} (\bibinfo {year} {2024})},\ \Eprint
  {http://arxiv.org/abs/2406.07321} {arXiv:2406.07321 [hep-ph]} \BibitemShut
  {NoStop}%
\bibitem [{\citenamefont {Chernyshev}\ \emph {et~al.}(2024)\citenamefont
  {Chernyshev}, \citenamefont {Robin},\ and\ \citenamefont
  {Savage}}]{Chernyshev:2024pqy}%
  \BibitemOpen
  \bibfield  {author} {\bibinfo {author} {\bibfnamefont {I.}~\bibnamefont
  {Chernyshev}}, \bibinfo {author} {\bibfnamefont {C.~E.~P.}\ \bibnamefont
  {Robin}}, \ and\ \bibinfo {author} {\bibfnamefont {M.~J.}\ \bibnamefont
  {Savage}},\ }\href@noop {} {\enquote {\bibinfo {title} {{Quantum Magic and
  Computational Complexity in the Neutrino Sector}},}\ } (\bibinfo {year}
  {2024}),\ \Eprint {http://arxiv.org/abs/2411.04203} {arXiv:2411.04203
  [quant-ph]} \BibitemShut {NoStop}%
\bibitem [{\citenamefont {Liu}\ \emph {et~al.}(2025)\citenamefont {Liu},
  \citenamefont {Low},\ and\ \citenamefont {Yin}}]{Liu:2025qfl}%
  \BibitemOpen
  \bibfield  {author} {\bibinfo {author} {\bibfnamefont {Q.}~\bibnamefont
  {Liu}}, \bibinfo {author} {\bibfnamefont {I.}~\bibnamefont {Low}}, \ and\
  \bibinfo {author} {\bibfnamefont {Z.}~\bibnamefont {Yin}},\ }\href@noop {}
  {\enquote {\bibinfo {title} {{Quantum Magic in Quantum Electrodynamics}},}\ }
  (\bibinfo {year} {2025}),\ \Eprint {http://arxiv.org/abs/2503.03098}
  {arXiv:2503.03098 [quant-ph]} \BibitemShut {NoStop}%
\bibitem [{\citenamefont {{CMS Collaboration}}(2025)}]{CMS-PAS-TOP-25-001}%
  \BibitemOpen
  \bibfield  {author} {\bibinfo {author} {\bibnamefont {{CMS Collaboration}}},\
  }\href {https://cds.cern.ch/record/2926751} {\emph {\bibinfo {title}
  {{Observation of magic states of top quark pairs produced in proton-proton
  collisions at $\sqrt{s}=13~\mathrm{TeV}$}}}},\ \bibinfo {type} {Tech. Rep.}\
  (\bibinfo  {institution} {CERN},\ \bibinfo {address} {Geneva},\ \bibinfo
  {year} {2025})\BibitemShut {NoStop}%
\bibitem [{\citenamefont {Aoude}\ \emph {et~al.}(2025)\citenamefont {Aoude},
  \citenamefont {Banks}, \citenamefont {White},\ and\ \citenamefont
  {White}}]{Aoude:2025jzc}%
  \BibitemOpen
  \bibfield  {author} {\bibinfo {author} {\bibfnamefont {R.}~\bibnamefont
  {Aoude}}, \bibinfo {author} {\bibfnamefont {H.}~\bibnamefont {Banks}},
  \bibinfo {author} {\bibfnamefont {C.~D.}\ \bibnamefont {White}}, \ and\
  \bibinfo {author} {\bibfnamefont {M.~J.}\ \bibnamefont {White}},\ }\href@noop
  {} {\  (\bibinfo {year} {2025})},\ \Eprint {http://arxiv.org/abs/2505.12522}
  {arXiv:2505.12522 [hep-ph]} \BibitemShut {NoStop}%
\bibitem [{\citenamefont {White}\ \emph {et~al.}(2021)\citenamefont {White},
  \citenamefont {Cao},\ and\ \citenamefont {Swingle}}]{White:2020zoz}%
  \BibitemOpen
  \bibfield  {author} {\bibinfo {author} {\bibfnamefont {C.~D.}\ \bibnamefont
  {White}}, \bibinfo {author} {\bibfnamefont {C.}~\bibnamefont {Cao}}, \ and\
  \bibinfo {author} {\bibfnamefont {B.}~\bibnamefont {Swingle}},\ }\href
  {\doibase 10.1103/PhysRevB.103.075145} {\bibfield  {journal} {\bibinfo
  {journal} {Phys. Rev. B}\ }\textbf {\bibinfo {volume} {103}},\ \bibinfo
  {pages} {075145} (\bibinfo {year} {2021})},\ \Eprint
  {http://arxiv.org/abs/2007.01303} {arXiv:2007.01303 [quant-ph]} \BibitemShut
  {NoStop}%
\bibitem [{\citenamefont {García}\ \emph {et~al.}(2014)\citenamefont
  {García}, \citenamefont {Markov},\ and\ \citenamefont
  {Cross}}]{Garcia:2014hia}%
  \BibitemOpen
  \bibfield  {author} {\bibinfo {author} {\bibfnamefont {H.}~\bibnamefont
  {García}}, \bibinfo {author} {\bibfnamefont {I.}~\bibnamefont {Markov}}, \
  and\ \bibinfo {author} {\bibfnamefont {A.}~\bibnamefont {Cross}},\ }\href
  {\doibase 10.26421/QIC14.7-8-9} {\bibfield  {journal} {\bibinfo  {journal}
  {Quantum Information and Computation}\ }\textbf {\bibinfo {volume} {14}},\
  \bibinfo {pages} {683} (\bibinfo {year} {2014})}\BibitemShut {NoStop}%
\bibitem [{\citenamefont {Gottesman}(1998)}]{Gottesman:1998hu}%
  \BibitemOpen
  \bibfield  {author} {\bibinfo {author} {\bibfnamefont {D.}~\bibnamefont
  {Gottesman}},\ }in\ \href@noop {} {\emph {\bibinfo {booktitle} {{22nd
  International Colloquium on Group Theoretical Methods in Physics}}}}\
  (\bibinfo {year} {1998})\ pp.\ \bibinfo {pages} {32--43},\ \Eprint
  {http://arxiv.org/abs/quant-ph/9807006} {arXiv:quant-ph/9807006} \BibitemShut
  {NoStop}%
\bibitem [{\citenamefont {Leone}\ \emph {et~al.}(2022)\citenamefont {Leone},
  \citenamefont {Oliviero},\ and\ \citenamefont {Hamma}}]{Leone:2021rzd}%
  \BibitemOpen
  \bibfield  {author} {\bibinfo {author} {\bibfnamefont {L.}~\bibnamefont
  {Leone}}, \bibinfo {author} {\bibfnamefont {S.~F.~E.}\ \bibnamefont
  {Oliviero}}, \ and\ \bibinfo {author} {\bibfnamefont {A.}~\bibnamefont
  {Hamma}},\ }\href {\doibase 10.1103/PhysRevLett.128.050402} {\bibfield
  {journal} {\bibinfo  {journal} {Phys. Rev. Lett.}\ }\textbf {\bibinfo
  {volume} {128}},\ \bibinfo {pages} {050402} (\bibinfo {year} {2022})},\
  \Eprint {http://arxiv.org/abs/2106.12587} {arXiv:2106.12587 [quant-ph]}
  \BibitemShut {NoStop}%
\bibitem [{\citenamefont {Wang}\ and\ \citenamefont {Li}(2023)}]{Wang:2023uog}%
  \BibitemOpen
  \bibfield  {author} {\bibinfo {author} {\bibfnamefont {Y.}~\bibnamefont
  {Wang}}\ and\ \bibinfo {author} {\bibfnamefont {Y.}~\bibnamefont {Li}},\
  }\href {\doibase 10.1007/s11128-023-04186-9} {\bibfield  {journal} {\bibinfo
  {journal} {Quant. Inf. Proc.}\ }\textbf {\bibinfo {volume} {22}},\ \bibinfo
  {pages} {444} (\bibinfo {year} {2023})}\BibitemShut {NoStop}%
\bibitem [{\citenamefont {Bhupal~Dev}\ and\ \citenamefont
  {Pilaftsis}(2014)}]{BhupalDev:2014bir}%
  \BibitemOpen
  \bibfield  {author} {\bibinfo {author} {\bibfnamefont {P.~S.}\ \bibnamefont
  {Bhupal~Dev}}\ and\ \bibinfo {author} {\bibfnamefont {A.}~\bibnamefont
  {Pilaftsis}},\ }\href {\doibase 10.1007/JHEP12(2014)024} {\bibfield
  {journal} {\bibinfo  {journal} {JHEP}\ }\textbf {\bibinfo {volume} {12}},\
  \bibinfo {pages} {024} (\bibinfo {year} {2014})},\ \bibinfo {note} {[Erratum:
  JHEP 11, 147 (2015)]},\ \Eprint {http://arxiv.org/abs/1408.3405}
  {arXiv:1408.3405 [hep-ph]} \BibitemShut {NoStop}%
\bibitem [{\citenamefont {Darvishi}\ \emph {et~al.}(2021)\citenamefont
  {Darvishi}, \citenamefont {Masouminia},\ and\ \citenamefont
  {Pilaftsis}}]{Darvishi:2021txa}%
  \BibitemOpen
  \bibfield  {author} {\bibinfo {author} {\bibfnamefont {N.}~\bibnamefont
  {Darvishi}}, \bibinfo {author} {\bibfnamefont {M.~R.}\ \bibnamefont
  {Masouminia}}, \ and\ \bibinfo {author} {\bibfnamefont {A.}~\bibnamefont
  {Pilaftsis}},\ }\href {\doibase 10.1103/PhysRevD.104.115017} {\bibfield
  {journal} {\bibinfo  {journal} {Phys. Rev. D}\ }\textbf {\bibinfo {volume}
  {104}},\ \bibinfo {pages} {115017} (\bibinfo {year} {2021})},\ \Eprint
  {http://arxiv.org/abs/2106.03159} {arXiv:2106.03159 [hep-ph]} \BibitemShut
  {NoStop}%
\bibitem [{\citenamefont {Darvishi}\ and\ \citenamefont
  {Pilaftsis}(2020)}]{Darvishi:2019dbh}%
  \BibitemOpen
  \bibfield  {author} {\bibinfo {author} {\bibfnamefont {N.}~\bibnamefont
  {Darvishi}}\ and\ \bibinfo {author} {\bibfnamefont {A.}~\bibnamefont
  {Pilaftsis}},\ }\href {\doibase 10.1103/PhysRevD.101.095008} {\bibfield
  {journal} {\bibinfo  {journal} {Phys. Rev. D}\ }\textbf {\bibinfo {volume}
  {101}},\ \bibinfo {pages} {095008} (\bibinfo {year} {2020})},\ \Eprint
  {http://arxiv.org/abs/1912.00887} {arXiv:1912.00887 [hep-ph]} \BibitemShut
  {NoStop}%
\bibitem [{\citenamefont {Seki}\ \emph {et~al.}(2015)\citenamefont {Seki},
  \citenamefont {Park},\ and\ \citenamefont {Sin}}]{Seki:2014cgq}%
  \BibitemOpen
  \bibfield  {author} {\bibinfo {author} {\bibfnamefont {S.}~\bibnamefont
  {Seki}}, \bibinfo {author} {\bibfnamefont {I.~Y.}\ \bibnamefont {Park}}, \
  and\ \bibinfo {author} {\bibfnamefont {S.-J.}\ \bibnamefont {Sin}},\ }\href
  {\doibase 10.1016/j.physletb.2015.02.028} {\bibfield  {journal} {\bibinfo
  {journal} {Phys. Lett. B}\ }\textbf {\bibinfo {volume} {743}},\ \bibinfo
  {pages} {147} (\bibinfo {year} {2015})},\ \Eprint
  {http://arxiv.org/abs/1412.7894} {arXiv:1412.7894 [hep-th]} \BibitemShut
  {NoStop}%
\bibitem [{\citenamefont {Darvishi}\ \emph {et~al.}(2024)\citenamefont
  {Darvishi}, \citenamefont {Pilaftsis},\ and\ \citenamefont
  {Yu}}]{Darvishi:2023fjh}%
  \BibitemOpen
  \bibfield  {author} {\bibinfo {author} {\bibfnamefont {N.}~\bibnamefont
  {Darvishi}}, \bibinfo {author} {\bibfnamefont {A.}~\bibnamefont {Pilaftsis}},
  \ and\ \bibinfo {author} {\bibfnamefont {J.-H.}\ \bibnamefont {Yu}},\ }\href
  {\doibase 10.1007/JHEP05(2024)233} {\bibfield  {journal} {\bibinfo  {journal}
  {JHEP}\ }\textbf {\bibinfo {volume} {05}},\ \bibinfo {pages} {233} (\bibinfo
  {year} {2024})},\ \Eprint {http://arxiv.org/abs/2312.00882} {arXiv:2312.00882
  [hep-ph]} \BibitemShut {NoStop}%
\end{thebibliography}%

\end{document}